\documentclass[aps,prr,reprint,longbibliography]{revtex4-2}
\usepackage{amsmath,amssymb}
\usepackage{graphicx}
\usepackage{dcolumn}
\usepackage{bm}
\newcommand{\vs}{\vspace{0.5em}}
\newcommand{\vsm}{\vspace{-0.5em}}
\newcommand{\ket}[1]{\ensuremath{\left|#1\right>}}
\newcommand{\bra}[1]{\ensuremath{\left<#1\right|}}

\newcommand{\mf}[1]{\boldsymbol{#1}}
\newcommand{\mc}[1]{\ensuremath{\mathcal{#1}}}

\newcommand{\matelal}[3]{\ensuremath{\left<#1\right.|#2\left.|#3\right>}}
\newcommand{\be}[1]{\begin{equation}#1\end{equation}}

\newcommand{\bal}[1]{\begin{align}#1\end{align}}

\begin{document}


\title{Probing quantum-coherent dynamics with free electrons}

\author{H. B. Crispin }
\email[Contact author: ]{crispin@physik.uni-kiel.de}
\affiliation{%
Institute for Experimental and Applied Physics, \\Christian Albrechts University, Leibnizstrasse 19, 24118 Kiel, Germany
}%
\author{N. Talebi}
\email[Contact author: ]{talebi@physik.uni-kiel.de}
\affiliation{%
Institute for Experimental and Applied Physics, \\Christian Albrechts University, Leibnizstrasse 19, 24118 Kiel, Germany
}%
\setcounter{footnote}{0}

\begin{abstract}
 Recent advances in time-resolved cathodoluminescence have enabled ultrafast studies of single emitters in quantum materials with femtosecond temporal resolution. Here, we
develop a quantum theory modeling the dynamics of free electrons
interacting with quantum emitters in arbitrary initial states. Our
analysis reveals that a free electron can induce transient coherent
oscillations in the populations when the system is initially prepared in a coherent superposition of its states. 
Moreover, the electron energy spectrum exhibits a clear signature of
the
quantum coherence and sensitivity to the transition frequency of the emitter. 
These coherence effects manifest themselves as oscillations in the zero-loss peak of the spectral energy-loss probability. Our findings pave the way for characterization of quantum-coherent dynamics of individual quantum emitters by electron-probes.     
\end{abstract}
\maketitle
\section{Introduction}
Free electrons are powerful tools for characterizing and imaging material excitations at the nanoscale \cite{Talebi2019a, *Talebi2017, Polman2019}. 
Spectroscopic techniques such as cathodoluminescence (CL) and electron-energy-loss spectroscopy (EELS) can probe optical transitions of defect states with high spatial resolution \cite{Hayee2020}. 
EELS also provides access to the projected local density of photonic states \cite{Abajo2008}. However, these measurements have not demonstrated sensitivity to the quantum coherence, i.e., coherent superposition of energy states. Solid-state defects act as quantum emitters and offer novel functionalities that arise from their quantum-coherent properties \cite{Heinrich2021}, making them central to modern quantum technologies \cite{Aharonovich2016, Awschalom2018, Loss1998}. Therefore, characterization of the coherence dynamics is of fundamental importance for free-electron-based spectroscopies.  

Several theoretical schemes have shown that shaped electron wavefunctions can probe the atomic coherence \cite{Gover2020, Ruimy2021, Zhao2021, Zhang2021, Zhang2022, Ruimy2023}. Therefore, the off-diagonal density-matrix elements of the system are accessible at a given time. 
However, in order to probe the emitter dynamics, repeated measurements must ensure that the target of interest, e.g., a defect structure, is not altered or damaged during acquisition. This poses a challenge in free-electron-matter interactions \cite{Roccapriore2024}. More importantly, there is considerable interest in resolving the quantum dynamics in materials \cite{Roccapriore2024, Lee2025, Taleb2025, Timmer2023}. Quantum systems evolve on ultrashort timescales due to strong interactions under ambient conditions \cite{Taleb2025, Timmer2023}. 
For example, defect centers in novel two-dimensional materials such as hBN exhibit strong coupling to phonons. In such cases, the radiative decay and dephasing time of the quantum emitters is of the order of hundreds of femtoseconds \cite{Taleb2025}. 
Furthermore, recent time-resolved techniques \cite{Taleb2023} demonstrate the ability to probe ultrafast quantum-coherent processes in materials, inspiring studies of the dynamics of single emitters in their natural environment \cite{Taleb2025}. This motivates a general theoretical description for the free-electron-emitter interactions.   

In this work, we develop a fully quantum, time-dependent theory for the coherent interaction between a free electron and quantum emitters in arbitrary initial states. The emitter is modeled as a two-level system, whose transition dipole moment couples to the Coulomb field of a moving Gaussian electron wavepacket. For various initial electron energies, we find that transient coherent oscillations can be induced in the populations when the system is prepared in an initial coherent superposition. In addition, we show that the electron energy spectrum is sensitive to the initial coherence and the transition frequency of the quantum emitter. These coherent effects appear as temporal oscillations centered on the zero-loss peak of the EELS probabilities. We further analyze the dependence of the EELS spectra on the relative phase of the constituents of the superposition. Finally, we connect our theoretical predictions to the relevant experimentally accessible observables and identify signatures of coherence in ultrafast CL, demonstrating the ability of free electrons to probe the quantum-coherent dynamics. \vsm \vsm

\section{Theoretical model \vsm} We consider the interaction between the emitter and the free electron in the configuration shown in Fig.~1(a). We assume the electron wavepacket propagates along the $z$-axis in the $x-z$ plane with velocity $v_{0}$. The quantum emitter, modeled as a two-level system with transition frequency $\omega_{0}$, is stationary at the origin. In this setup, the coherent interaction between the free electron and the emitter is given by the Hamiltonian \cite{Ruimy2021}
\bal{
H=H_{\textrm{qe}}+H_{\textrm{el}}+(\mf{d}_{eg} S^{+}+\mf{d}^{*}_{eg} S^{-})\cdot \vec{E}(r_{\perp},0,z),\label{hamil}\vs 
}
\begin{figure*}[t]
    \centering
     \includegraphics[width=0.99\textwidth, height=7.5cm]{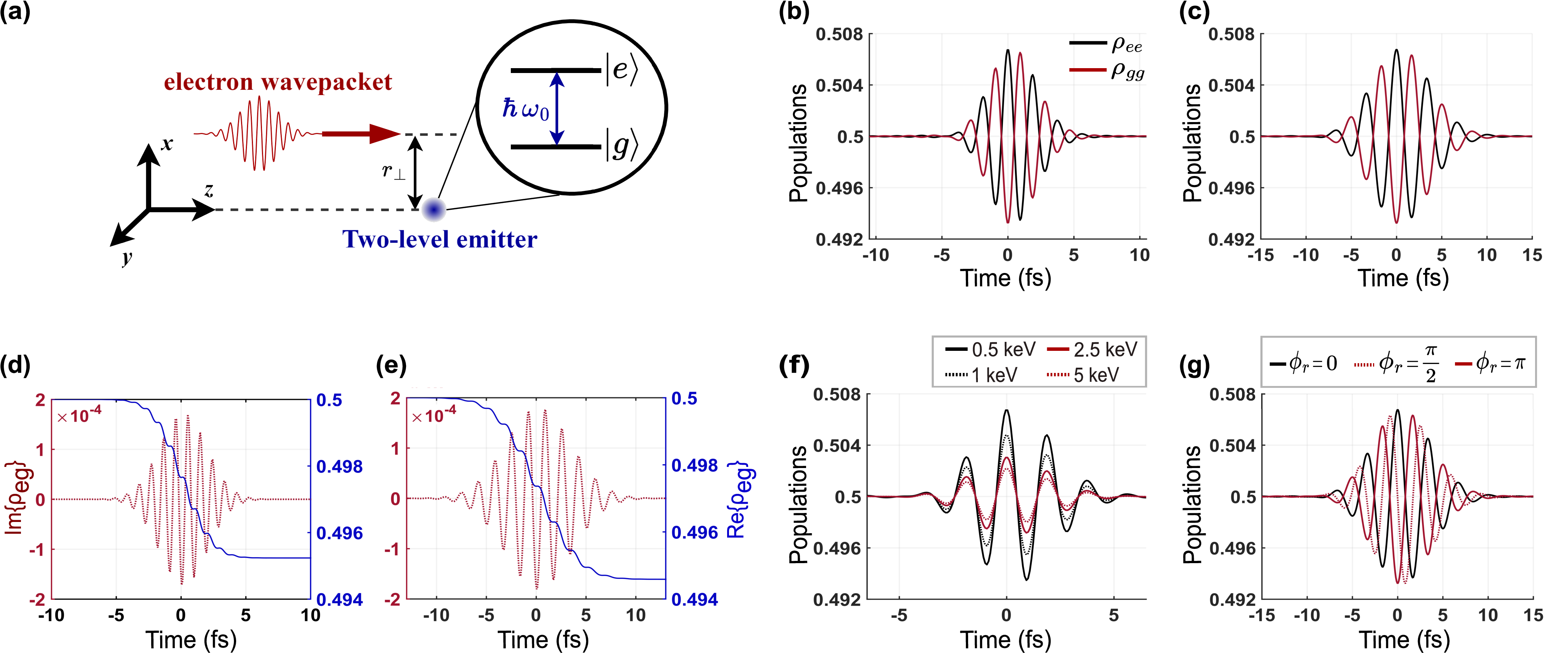}
\caption{(a) Schematic of the interaction between a two-level emitter and an electron wave packet. (b)–(e) Population (b),(c) and coherence (d),(e) dynamics for a $0.5$ keV electron with impact parameter $r_{\perp}=2$ nm. The emitter has $\mf d_x=\mf d_z=30$ Debye and transition wavelengths of 560 nm in (b),(d) and 1000 nm in (c),(e); the initial state is $(\ket g+\ket e)/\sqrt{2}$. (f) Effect of increasing electron energy on the population dynamics for a 560 nm transition. (g) Effect of initial coherence on the population dynamics for a 1000 nm transition.}\vsm \vsm \vsm 
\end{figure*}where $H_{\textrm{qe}}=\hbar \omega_{0} S_{z}$ and $H_{\textrm{el}}=-i \hbar v_{0} \partial_{z}$ are the free Hamiltonians of the quantum emitter and the propagating electron, respectively. Here $S_{z}=(\ket{e}\bra{e}-\ket{g}\bra{g})/2$, and the transition operators are defined as $S^{+}=\ket{e}\bra{g}$ and $S^{-}=\ket{g}\bra{e}$. The third term describes the interaction of the electron with the transition dipole moment $\mf{d}_{eg}=\mf{d}_{x} \hat{e}_{x}+\mf{d}_{y} \hat{e}_{y}+\mf{d}_{z} \hat{e}_{z}$ of the two-level system via its Coulomb field \cite{feynman1964},
\bal{
\vec{E}(r_{\perp},0,z)=\frac{-e \gamma}{4 \pi \epsilon_{0}}\frac{(z \hat{e}_{z}+r_{\perp} \hat{e}_{x})}{(\gamma^{2}z^{2}+r_{\perp}^{2})^{3/2}}.\vs
}
Here, $\gamma=1/
\sqrt{1-v_{0}^2/c^2}$ is the Lorentz factor, $\epsilon_{0}$ is the free space permittivity, and $r_{\perp}$ is the impact parameter that represents the distance between the center of the moving electron wavepacket and the two-level emitter. 

In obtaining the Hamiltonian (\ref{hamil}), we employed the non-recoil approximation for the free electron.
This approximation is justified when the initial electron energy $E_{0}$ is much greater than the transition energy $\hbar \omega_{0}$. 
We further assume that the interaction timescales involved are typically much shorter than the decoherence and dephasing times.  We therefore neglect these effects in the present treatment. 

To describe the dynamics of the total system including the free electron and the quantum emitter, it is helpful to move to the interaction picture $H_{I}(t)=U (H-H_{0})U^{\dagger}$ with respect to the unitary transformation $U=\exp{\{i H_{0}t/\hbar\}}$, where $H_{0}=H_{\textrm{qe}}+H_{\textrm{el}}$. The formal solution is
\bal{
\ket{\psi(t)}= \mc{T} e^{-\frac{i}{\hbar}\int^{t}_{-\infty} H_{I}(t_{1}) dt_{1}}\ket{\psi_{in}},
}
where \mc{T} denotes time ordering and $\ket{\psi_{in}}=\ket{\psi_{qe}}\ket{\psi_{el}}$ is the initial state. We consider the quantum state of the two-level emitter to be in a general coherent superposition $\ket{\psi_{qe}}=a\ket{g}+b\, e^{-i \phi_{r}}\ket{e}$ with $\phi_{r}$ being the relative phase. The initial free-electron state is a Gaussian wavepacket in the position basis,
\bal{
\ket{\psi_{el}}= (2\pi \sigma_{z}^2)^{-1/4}\int dz\, e^{-\frac{z^2}{4 \sigma_{z}^2}+i k_{0}z}\ket{z},
}
where $k_{0}=E_{0}v_{0}/\hbar c^2$ for an electron of initial energy $E_{0}$ and  spatial width $\sigma_z$. 

Using the Magnus expansion \cite{Magnus1954}, the joint free electron-emitter state, upto a normalization factor, can be written in a compact form
\be{
\ket{\psi(t)} = \int dz\, e^{-\frac{z^2}{4\sigma_{z}^2}+i k_{0} z} [c_{g}(z,t) \ket{g}+c_{e}(z,t)\ket{e}]\ket{z},\label{main}
}
with probability amplitudes,
\bal{
c_{g}(z,t)=&~\big[\,a \cos{|g(z,t)|}-i\, b \,e^{-i( \phi_{r}+\phi_{g}(z,t))}\sin{|g(z,t)|}\,\big],&\nonumber\\
c_{e}(z,t)=&~\big[\,b\, e^{-i \phi_{r}} \cos{|g(z,t)|}-i\, a\, e^{i\phi_{g}(z,t)}\sin{|g(z,t)|}\,\big].\label{probamp}
}
The expressions for the amplitudes $c_{g}(z,t)$ and $c_{e}(z,t)$ clearly show a relative phase dependence. In deriving (\ref{main}), the higher-order terms in the Magnus expansion have been neglected, which is justified due to the weak coupling, $|g(z,t)|\ll1$. In contrast to previous studies \cite{Ruimy2021, Zhao2021}, the coherent interaction is now governed by a complex,\textit{ time-dependent} coupling strength $g(z,t)=|g(z,t)|e^{i\phi_{g}(z,t)}$, where $\phi_{g}(z,t)$ is the phase (Eq.~(\ref{ganal}) in Appendix A). \vsm
\begin{figure*}[t]
    \centering
     \includegraphics[width=0.97\textwidth]{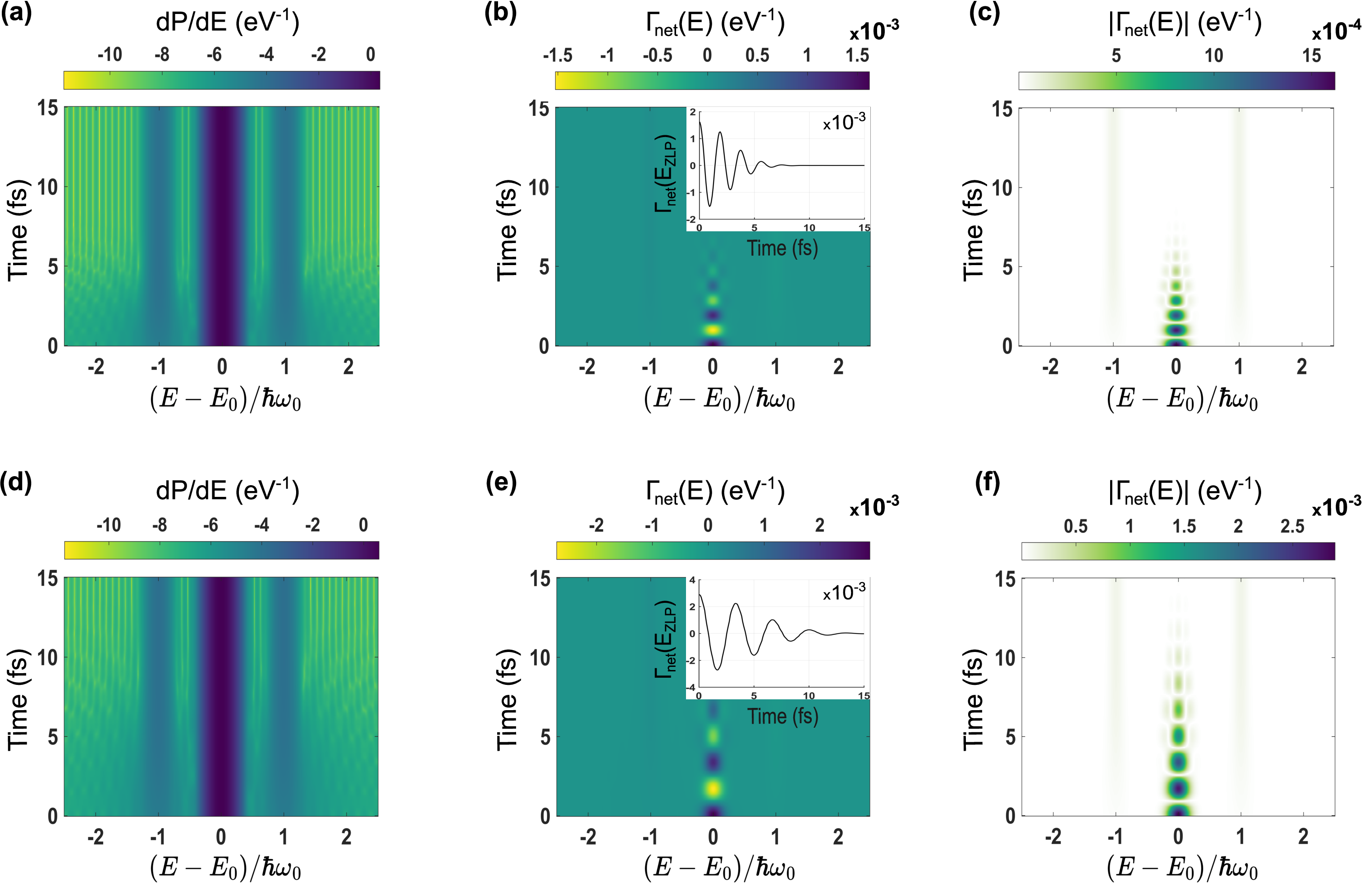}
\caption{(a),(d) Time evolution of the EELS probability $dP/dE$ shown on a logarithmic scale for an electron with kinetic energy $\simeq 115$ keV ($v_{0}=0.578c$) for $t\ge0$ and impact parameter $r_{\perp}=2$nm. The two-level emitter parameters in (a)-(c) and (d)-(f) are the same as in Fig.~1(b) and Fig.~1(c), respectively. The corresponding EELS probability differences $\Gamma_{net}(E)$ in (b),(e) and their absolute values $|\Gamma_{net}(E)|$ in (c), (f), show coherent oscillations in the zero-loss peak for the same electron velocity and impact parameter as in (a),(d).}\vsm \vsm 
\end{figure*}

\section{Dynamics of the emitter \vsm} We first consider the evolution of the two-level emitter, which can be obtained from the reduced density matrix $\rho_{qe}(t)=\textrm{Tr}_{el}\{\rho(t)\}$ with $\rho(t)=\ket{\psi(t)}\bra{\psi(t)}$. Figs.~1(b) and 1(c) depict the population dynamics for quantum emitters whose transition frequencies lie in the optical ($\hbar \omega_{0} \simeq 2.2$ eV) and the near-infrared region ($\hbar \omega_{0} \simeq 1.2$ eV), respectively. The corresponding evolution of the coherences are displayed in Figs.~1(d) and 1(e). The free electron induces \textit{Rabi-like} transient coherent oscillations in the populations and coherences when the electron wavepacket duration $\sigma_{t}=\sigma_{z}/v_{0}$ is comparable to or larger than the optical period $T=2\pi/\omega_{0}$. The role of the electron wavepacket duration $\sigma_{t}$, including the regimes $\sigma_{t}\sim T$ and $\sigma_{t}>T$, is discussed in Sec.~\textbf{V}. The temporal envelope of these oscillations varies as a Gaussian $\sim e^{-t^2/\sigma_{t}^2}$. As a result, in comparison to optical transitions, one can see that the coherence effects occur over comparatively longer timescales for quantum emitters with infrared transitions [compare Fig.~1(b) and 1(c)]. 
We emphasize that the observed oscillatory features are prominent for coherent excitations by low energy electrons [Fig.~1(b)-(c) and 1(f)]. In comparison, the dynamics are difficult to be observed for fast electrons $E>30$keV. The effect of varying initial electron energies on the populations is shown in Fig.~1(f). For increasing electron velocities, we find that the coherent oscillations are suppressed. Nevertheless, the temporal features are still visible even for energies of a few keVs. More importantly, the population dynamics are sensitive to the relative phase $\phi_r$ of the initial superposition state. This dependence is illustrated in Fig.~1(g) for an equally weighted coherent superposition $\ket{\psi_{in}}=(\ket{g}+e^{-i \phi_{r}}\ket{e})/\sqrt{2}$. A relative phase change from $\phi_{r}=0$ to $\phi_{r}=\pi$ acts as a control parameter over the oscillations induced by the electron wavepacket. Therefore, it is interesting to investigate the electron energy spectrum for a signature of the quantum coherence of the two-level emitter. \vsm \vsm \vsm 

\section{EELS dynamics\vsm \vsm} To explore the coherence effects in the spectral electron energy loss probability, we compute the reduced density matrix of the free electron $\rho_{el}(t)=\textrm{Tr}_{qe}\{\rho(t)\}$. The diagonal elements, $dP/dE=\matelal{k(E)}{\rho_{el}(t)}{k(E)}$, in $k-$space provide information about the EELS,
\bal{
\frac{dP}{dE}=\frac{1}{\hbar v_{0}}\sum_{i=g,e}|\psi_{i}(k(E), t)|^2,\label{eels}\vsm \vsm 
}
where $\psi_{i}(k(E),t)= N^{-1}\!\int \! dz \,e^{-z^2/4 \sigma_{z}^2}e^{-i k z}c_{i}(z,t)$, $N=\sqrt{2\pi}\,(2\pi\sigma_z^2)^{1/4}$ is a multiplicative factor, and $k(E)=(E-E_{0})/\hbar v_{0}$. Notably, the spectrum (\ref{eels}) is time-dependent and can be viewed as a time-resolved electron energy-loss spectrum. The terms $\psi_{g}$ and $\psi_{e}$ in Eq.~(\ref{eels}) can be understood as the spectral probability amplitudes associated with the electron gaining and losing energy due to the interaction with the two-level emitter, respectively. The sum of the absolute square of these individual amplitudes gives the EELS probabilities of electron energy-loss $|\psi_{e}|^2$ and gain $|\psi_{g}|^2$. In the postinteraction EELS ($t\rightarrow \infty$), the quantum coherence information is lost for excitation by unshaped electrons \cite{Ruimy2021}. However, the dynamics of the electron energy spectra reveals novel coherent properties. 

We are interested in times $t>0$ for which the electron wavepacket has crossed the two-level emitter. The spectral energy-loss probability for transition energies $\hbar \omega_{0} \simeq 2.2$ eV and $\hbar \omega_{0} \simeq 1.2$ eV can be seen in Figs.~2(a) and 2(d), respectively for $t\ge0$. 
The electron energy gain $E=E_{0}+\hbar \omega_{0}$ and loss peaks $E=E_{0}-\hbar \omega_{0}$ are clearly visible in the spectra along with the prominent zero-loss peak $E=E_{0}$. We note that the probabilities associated with electron energy loss $|\psi_{e}|^2$ and gain $|\psi_{g}|^2$ depend on the relative phase $\phi_{r}$ [from Eq.~(\ref{probamp}) and (\ref{eels})]. 

To clearly see the signature of quantum coherence in the EELS dynamics, we consider the probability difference
\bal{
\Gamma_{net}(E)=\frac{1}{\hbar v_{0}}\big(|\psi_{e}(E)|^2-|\psi_{g}(E)|^2 \big),
}
where the time-dependence is implied. A remarkable feature is the coherent oscillations in the zero-loss peak [Figs.~2(b) and 2(e)], signifying the dynamics of the energy exchange between the electron wavepacket and the emitter. The minimum and maximum probabilities at a given time correspond to the net energy gained or lost by the free electron to the quantum emitter. The zeroes in the spectra [Figs.~2(b) and 2(e)] imply a net zero energy exchange, i.e., they denote times for which the emitter returns to the same initial state $\ket{\psi_{in}}$. A contour of the absolute values $|\Gamma_{net}(E)|$ clearly depicts the timescales for which the induced coherence effects survive [Figs.~2(c) and 2(f)], which agrees well with the corresponding results in Fig.~1(b)-(c). For an initial electron velocity $v_{0}=0.578\, c$ ($E\simeq 115$keV), the distance traversed by the electron after crossing the emitter for times $t\sim 4-6$ fs is comparable to or greater than the wavelength of the transitions considered. The induced coherence effects can therefore play a significant role for collective excitations of nearby quantum emitters. 
Interestingly, the period of the oscillations observed in the zero-loss peak [Fig.~2(b) and (e)] is determined by the transition frequency $\omega_{0}$ of the quantum emitter being probed. In addition, the resulting dynamics is sensitive to the quantum coherence [Fig.~3]. The relative phase $\phi_{r}$ now acts as a tunable parameter to control the temporal features in the EELS spectra [Fig.~3(a)-(d)]. For shorter impact parameters, the probabilities oscillate with larger amplitudes [compare Fig.~3(a) and 3(b); 3(c) and 3(d)]. 
\begin{figure}[t]
    \centering
     \includegraphics[width=0.49\textwidth, height=7.5cm]{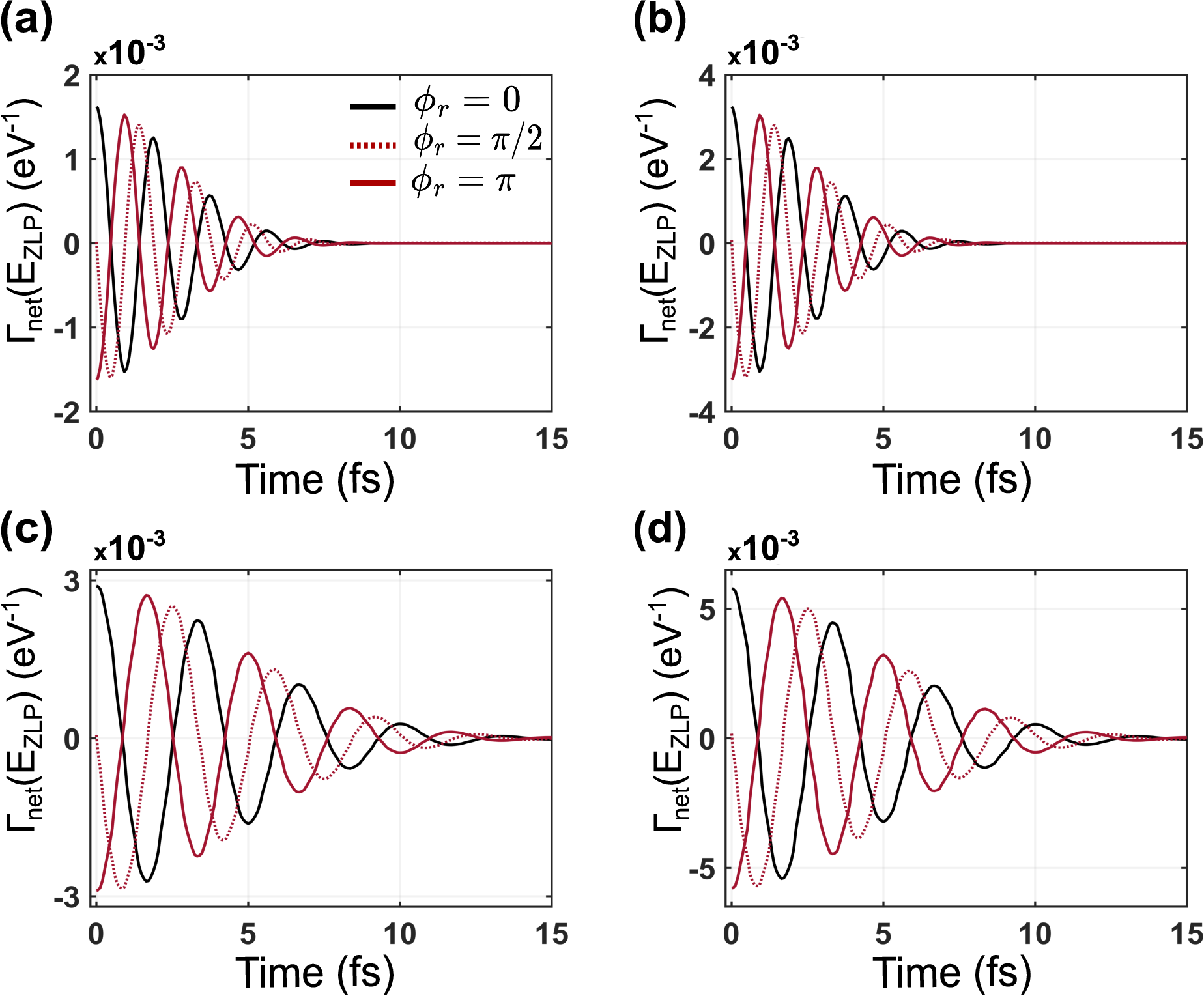}
\caption{(a)-(d) Dependence of the coherent oscillations in the zero-loss peak on the initial quantum coherence $(\ket{g}+e^{-i \phi_{r}}\ket{e})/\sqrt{2}$. The relative phase effects shown for two impact parameters (a),(c) $r_{\perp}=2$nm and (b),(d) $r_{\perp}=1$nm. The transition wavelengths used in (a),(b) and (c),(d) are the same as in Fig.~2(b) and Fig.~2(e), respectively. 
}
\end{figure}

To intuitively understand these results, it is useful to express $g(z,t)$ in the form, 
\be{
g(z,t)=~\frac{1}{2\pi v_{0}} \int \!\!d\omega\, \tilde{g}\bigg(\frac{\omega-\omega_{0}}{v_{0}}\bigg) e^{-i \frac{\omega}{v_{0}}z}e^{-i(\omega-\omega_{0})t}, \label{gfreq}
}
where the coupling parameter $\tilde{g}(\omega)$ can be written as the sum of two parts,
\bal{
\tilde{g}(\omega)= \big[\pi \delta\left(\omega\right) g_{\infty}+g_{c}(\omega)\big],
}
such that the oscillatory exponential factor is weighted by both a constant term $g_{\infty}$ and a frequency-dependent term $g_{c}(\omega)$, respectively (see Appendix A for derivations and explicit forms). It is clear that the Dirac delta contribution leads to exact energy translations $b(\omega_{0})\ket{E}=\ket{E-\hbar \omega_{0}}$ and $b^{\dagger}(\omega_{0})\ket{E}=\ket{E+\hbar \omega_{0}}$, where $b(\omega_{0})=e^{-i \omega_{0} z/v_{0}}$ can be identified as the electron momentum ladder operator $b(\omega_{0})\ket{E}=\ket{E-\hbar \omega_{0}}$. Therefore, the electron can gain or lose energy $\hbar \omega_{0}$ due to the interaction with the two-level emitter \cite{Ruimy2021}. A novel feature is the transient oscillatory component, which leads to intriguing phenomena in the dynamics of the EELS spectrum and the quantum emitter discussed here. In the long-time limit $t\rightarrow \infty$, we mention that the dynamical coupling reduces to the standard interaction constant $g$ \cite{Gover2020, Ruimy2021, Zhao2021} (see Appendix A for detailed discussion).

\section{Discussion and experimental feasibility} The aim of coherent spectroscopic techniques is to probe the coherence of material excitations and to characterize their quantum-coherent dynamics. A major challenge is to coherently control and manipulate the quantum states of emitters in their local environment. Free electrons provide the high spatial resolution needed to access the coherence of individual quantum emitters prepared in superposition states. Beyond the probing of static off-diagonal density-matrix elements that requires preshaped electrons \cite{Ruimy2021, Zhao2021}, the time evolution of quantum superpositions can be tracked even with unmodulated electron wavepackets. This opens the way to novel transient spectroscopic techniques offering high spatial and temporal resolution for the study of dynamical processes in quantum systems. 
\begin{figure}[t]
    \centering
     \includegraphics[width=5cm,height=10cm]{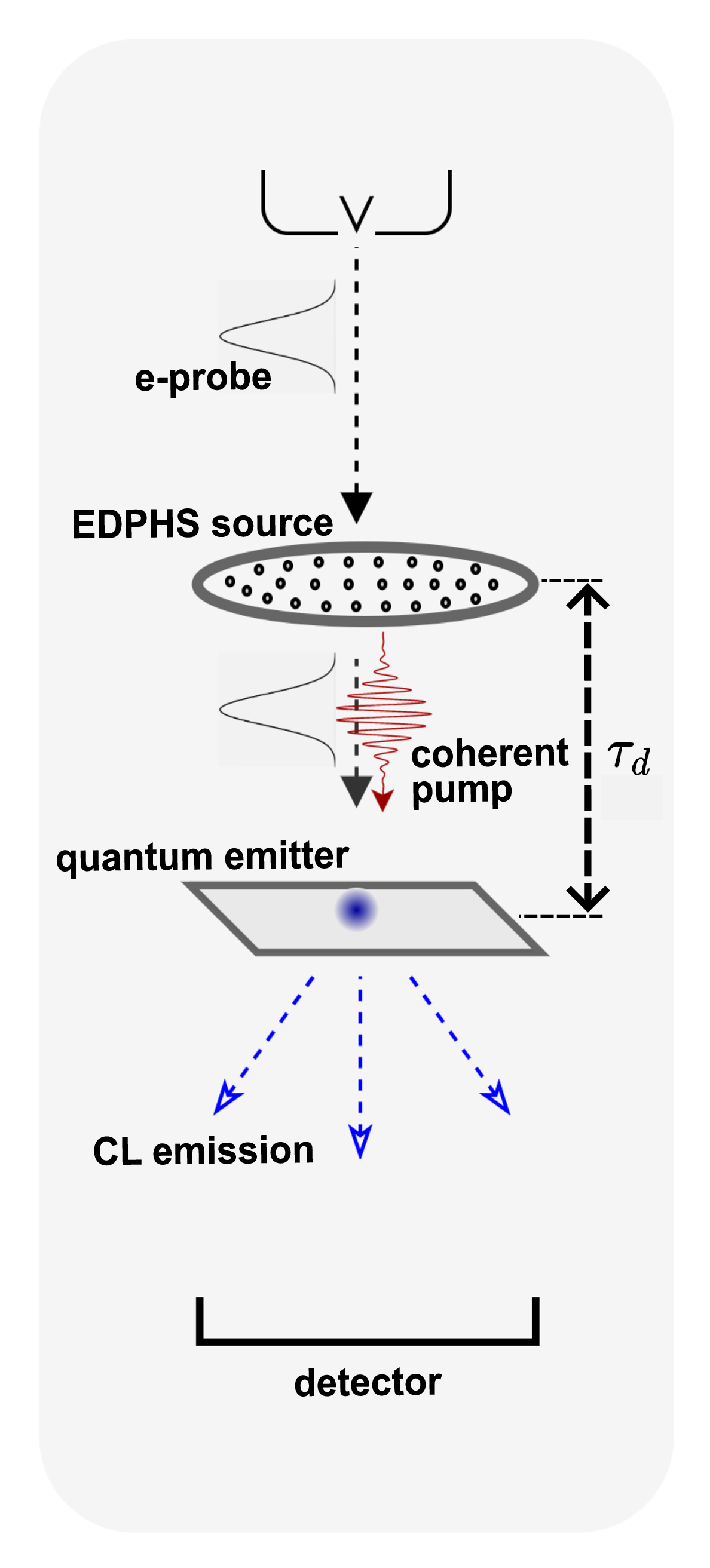}
\caption{ Experimental scheme of an ultrafast scanning electron microscope in which the electron beam strikes a sieve-like metal structure (EDPHS), generating focused coherent light pulses. The resulting coherent radiation creates a coherent superposition of states in the emitter, which is subsequently probed by the free electron. A controllable delay ($\tau_{d}$) enables access to the ultrafast dynamics of the quantum emitter.\label{edphsscheme} 
}
\end{figure}
\begin{figure}[t]
    \centering
     \includegraphics[width=0.38\textwidth]{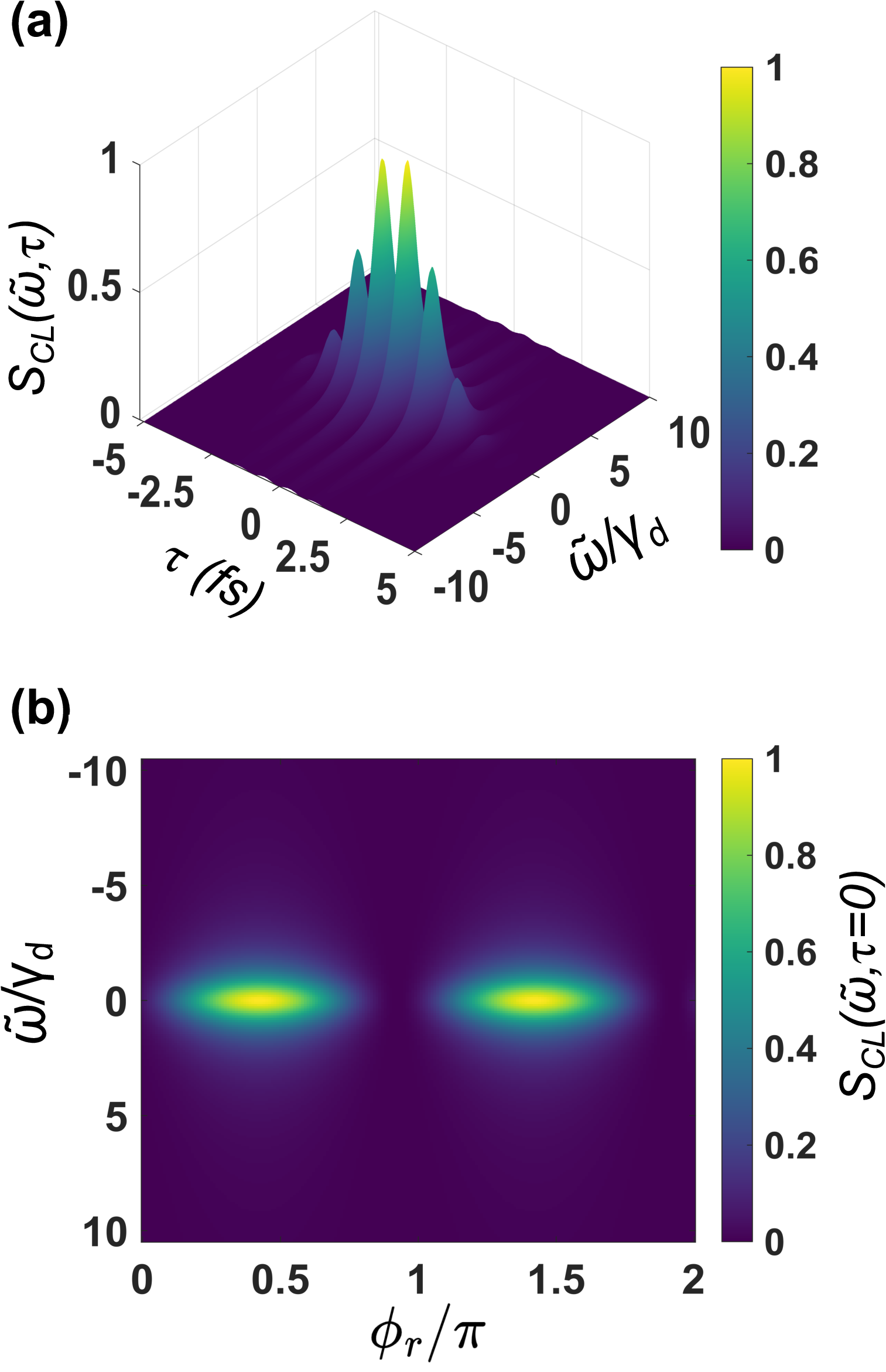}
\caption{The parameters used in the simulations of the normalized CL spectra $S_{CL}(\tilde{\omega}, \tau)$ are the same as in Fig.~1(b), with decoherence rate $\gamma_{d}=(\gamma_{r}+2 \gamma_{nr})/2$. The radiative decay and dephasing rates are $\gamma_{r}=10^{11}\,s^{-1}$ and $\gamma_{nr}=2 \times 10^{13} \,s^{-1}$, respectively. Panel (a) corresponds to a fixed initial coherent superposition state $(\ket{g}+\ket{e})/\sqrt{2}$ and (b) shows the spectra $S_{CL}(\tilde{\omega}, \tau=0)$ for different initial states $(\ket{g}+e^{-i \phi_{r}}\ket{e})/\sqrt{2}$. \label{ramseyspectra} 
}
\end{figure}

We now discuss the experimental feasibility of observing the theoretical predictions. The transient features in the population dynamics occur on femtosecond timescales. To probe such ultrafast coherent processes in time-resolved CL, electron-driven photon sources (EDPHS) are suitable candidates \cite{Taleb2023}. Spectroscopic techniques based on EDPHS have already demonstrated the ability to probe single quantum emitters with femtosecond time resolution \cite{Taleb2025}. 

In pump-probe schemes relying on free electrons [see Fig.~(\ref{edphsscheme})], the electron acts as a probe, while a coherent optical field serves as the pump that prepares the quantum emitter in an initial coherent superposition state. The pump is the radiation generated by the electron probe upon interaction with an electron-driven photon source \cite{Christopher2018, Talebi2019}. Mutual coherence between the optical field and the electron wavepacket provides a stable phase reference for successive electron--light interactions with the quantum emitter. The delay $\tau_d$ between the pump field and the electron probe can be precisely controlled by changing the distance between the EDPHS and the sample as the electron and photons travel at different speeds, thereby defining the probe arrival time and enabling access to the emitter's ultrafast dynamics. In addition, the delay $\tau_{d}$ sets the phase $\phi_{r}$ of the superposition state at the arrival time of the free electron. 

Recent experiments have demonstrated attosecond-to-femtosecond control of this relative delay $\tau_d$ \cite{Taleb2025}. As a result of this high temporal resolution, the coherent dynamics can be probed on timescales significantly shorter than those associated with decoherence and other competing inelastic relaxation channels. Therefore, decoherence effects from such secondary processes are strongly suppressed, and the coherent component of the emitted CL radiation dominates the measured signal. Consequently, for delay times $\tau_d \ll 1/\gamma_{d}$, i.e., much shorter than the decoherence and dephasing timescales, interference effects emerge in the CL emission, directly probing the coherent polarization of the system.

The corresponding experimental observable is thus governed by the induced coherence [see Appendix D] and is given by
\be{
S_{CL}(\tilde{\omega}, \tau)=\left|\int_{\tau}^{\infty}\!\!dt\, e^{i \tilde{\omega} t} \rho_{eg}(t,\tau)\right|^2, \label{ramspec}
}
where the spectrum is expressed as a function of the detuning $\tilde{\omega}=\omega-\omega_{0}$ from the transition frequency $\omega_{0}$, and of the electron probe arrival time $\tau$. Here, $\rho_{eg}(t,\tau)$ denotes the induced coherence in the presence of decay and dephasing in the interaction picture defined by $U=\exp\{i H_{0} t/\hbar\}$ (see Appendix D for the derivation). 

For a quantum emitter initially prepared in an equally weighted coherent superposition state ($a=b$), the spectrum admits the following analytical form:
\bal{
S_{CL}(\tilde{\omega}, \tau)=&\bigg|\frac{e^{-\gamma_{d}(\tau-t_{0})}
}{\gamma_{d}-i\, \tilde{\omega}} a\, b\, e^{- i \phi_{r}}\nonumber\\&+\frac{a b}{\pi \hbar v_{0}} \int \!\!d\Omega \,\frac{f_{el}(\Omega+\omega_{0})\, e^{-i\Omega\tau}}{\gamma_{d}-i \tilde{\omega}}\bigg|^2,\label{rmspec}
}
where $\tilde{\omega}=\omega-\omega_{0}$ with $\omega$ being the observed frequency, $t_{0}$ denotes the initial time of preparation of the coherent superposition state, and $\tau$ denotes the arrival time of the electron probe. The quantity $S_{CL}(\tilde{\omega},\tau)$ represents the coherent CL radiation. From the analytical expression, we see that the spectrum is a weighted Lorentzian that is centered at the frequency $\tilde{\omega}=0$, with a width set by the total decoherence rate $\gamma_{d}$ of the quantum emitter. The weight $f_{el}(\omega)$, which arises from the coherent interaction between the electron probe and the emitter, depends explicitly on the relative phase $\phi_r$ (see Appendix D). Consequently, the CL spectrum exhibits quantum interference effects. 
The numerical results are shown in Fig.~(5). A notable feature is the appearance of oscillatory peaks in the CL emission [Fig.~5(a)], analogous to Ramsey fringes, providing a clear signature of quantum interference in ultrafast CL. To further demonstrate the sensitivity of the spectrum to the initial coherence, we plot $S_{CL}(\tilde{\omega},\tau=0)$ for different relative phases $\phi_r\in[0,2\pi]$ [see Fig.~5(b)]. The resulting alternating maxima and minima in the CL intensity directly reveal the quantum coherence of the emitter and illustrate how coherent quantum dynamics can be probed using free electrons. We note that an equally weighted superposition of the emitter states is not essential for observing the interference features in CL, as follows from Eq.~(D15). The coherence effects persist even for superposition states with a larger ground-state population. 

\begin{figure}[t]
    \centering
     \includegraphics[width=0.45\textwidth]{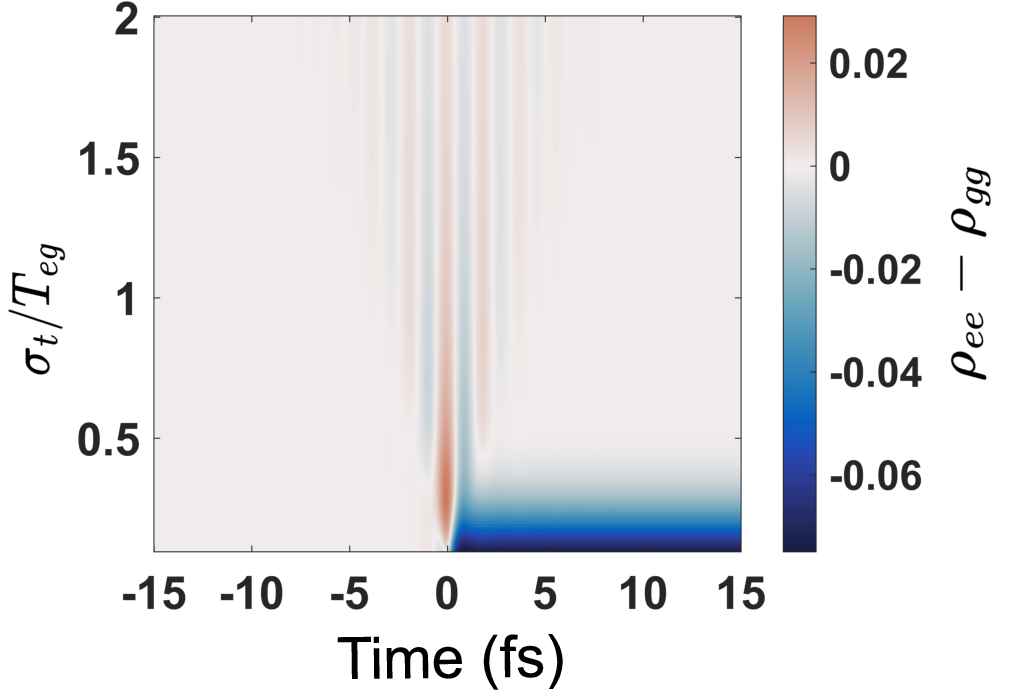}
\caption{Dynamics of the population inversion $\rho_{ee}-\rho_{gg}$ as a function of the electron wavepacket duration $\sigma_{t}$ for a $1$ keV electron with impact parameter $r_{\perp}=2$ nm. The emitter has $\mf d_x=\mf d_z=30$ Debye and transition wavelength of 560 nm with the initial state being $(\ket g+\ket e)/\sqrt{2}$.\label{popinv} 
}\vsm \vsm 
\end{figure}

Additionally, it is important to note that the relevant parameter regime [see Fig.~\ref{popinv}] for observing these quantum coherence effects corresponds to $\sigma_{t} > T_{eg} = 2\pi/\omega_{0}$, where $\omega_{0}$ denotes the transition frequency of the quantum emitter. In contrast, for $\sigma_{t} \ll T_{eg}$, the incident electron primarily induces population transfer between the two levels. The coherent oscillations emerge only when the free-electron wavepacket duration becomes comparable to the transition period $T_{eg}$. For $\sigma_{t}>T_{eg}$, we find that the induced coherent oscillations persist and occur over longer timescales but with decreasing amplitudes as the wavepacket duration $\sigma_{t}$ increases. Therefore, in the limit $\sigma_{t}\gg T_{eg}$, the coherence effects become suppressed while the oscillations are most prominent around $\sigma_{t} \sim T_{eg}$. To observe the dynamical coherence effects in EELS, a key challenge would be the implementation of novel \textit{in situ} measurements to track the evolution of the electron energy spectrum. Suitable model quantum systems can be 2D materials that host defect centers  \cite{Shaik2021, Azzam2021}. Additionally, quantum dots such as InGaAs/GaAs act as ideal two-level emitters \cite{Kamada2001, Stievater2001, Patton2005, Xu2007}. These materials exhibit dipole moments of the order of 20-40 debye \cite{Silverman2003} and transition frequencies in the IR region, which is within the range of parameters considered here.   

We next discuss the validity of the theoretical framework. The coupling regime explored here is for weak interactions, $\textrm{Max}\{|g(z,t)|\}\ll1$. For material systems with very large dipole moments such as perovskite nanocrystals \cite{Sutherland2016, Jena2019} and Rydberg atoms \cite{Hulet1985}, there can be an enhanced electron-emitter coupling resulting in stronger coherence effects. The higher-order terms in the Magnus expansion could be important in such cases. In addition, impact parameters smaller than the size of the emitter may result in corrections to the dipole approximation. 


\section{Summary} In conclusion, our analysis reveals the potential of free electrons to probe the coherence dynamics in quantum emitters. For initial superposition states, the populations exhibit transient oscillations that are sensitive to the relative phase. A crucial finding is the signature of quantum coherence of the emitter in the time-dependent EELS spectra and ultrafast cathodoluminescence. 
Looking forward, generalizations to multilevel quantum systems \cite{Crispin2025}, ensembles \cite{Ruimy2023}, and interactions with shaped electron wavefunctions \cite{Ruimy2021, Talebi2020, Chahshouri2023, Chahshouri2025} would enable the study of the excited-state dynamics and entanglement evolution in quantum systems. The theoretical framework therefore lays the ground for related works. Finally, the ability to probe quantum-coherent dynamics with free electrons is attractive for exploring ultrafast coherent-state transfer in quantum spin chains with high spatial and temporal resolution.   

\section{Acknowledgements} This project has received funding from the Volkswagen Foundation (Momentum Grant) and from the European Research Council (ERC) under the European Union’s Horizon 2020 research and innovation programme, under grant agreement No. 101170341 (ERC Consolidator Grant UltraSpecT) and grant agreement No. 101017720 (EBEAM).
\appendix
\section{Time-dependent quantum theory for the coherent interaction between a free electron and a quantum emitter}
In this section, we present a formulation of the theory describing the quantum-coherent dynamics in the populations and EELS. We start from the Hamiltonian that governs the coupling of the two-level emitter to the evanescent electric field of the propagating electron, 
\bal{
H=\hbar \omega_{0} S_{z}-i \hbar v_{0} \partial_{z}+(\mf{d}_{eg} S^{+}+\mf{d}^{*}_{eg} S^{-})\cdot \vec{E}(r_{\perp},0,z),\label{hamilt}\vs 
}
where $S_{z}=(\ket{e}\bra{e}-\ket{g}\bra{g})/2$ and the transition operators are denoted by $S^{+}=\ket{e}\bra{g}$ and $S^{-}=\ket{g}\bra{e}$. This model Hamiltonian has been extensively discussed in the context of coherently modulated electron wavepackets \cite{Gover2020, Ruimy2021, Zhao2021, Zhang2021, Zhang2022} and its derivation can be found in Ref.~\cite{Ruimy2021}. 
To probe the ultrafast quantum dynamics, we must consider the time-dependent evolution of the joint free-electron and emitter. This is relevant for quantum emitters in emerging two-dimensional material systems and essential for understanding the coherent interactions in recent time-resolved CL measurements \cite{Taleb2025}. 

As mentioned in the main text, we use the interaction picture $H_{I}(t)=U (H-H_{0})U^{\dagger}$ with respect to the unitary transformation $U=\exp{\{i H_{0}t/\hbar\}}$, where $H_{0}=\hbar \omega_{0} S_{z}-i \hbar v_{0} \partial_{z}$. The solution can be written as 
\bal{
\ket{\psi(t)}=e^{\sum_{k}\Omega_{k}(t)}\ket{\psi(t_{0})},\label{sol}
}
where $\ket{\psi(t_{0})}=\ket{\psi_{qe}}\ket{\psi_{el}}$ is the initial state of the free electron and the emitter in the interaction picture and $\Omega_{k}$ are the terms in the Magnus expansion. We describe the free electron by a Gaussian wavepacket in the position basis,
\bal{
\ket{\psi_{el}}= \frac{1}{(2\pi \sigma_{z}^2)^{1/4}}\int dz\, e^{-\frac{z^2}{4 \sigma_{z}^2}+i k_{0}z}\ket{z}.\label{elecstat}
}
Here, $k_{0}=E_{0}v_{0}/\hbar c^2$ for an initial electron energy $E_{0}=c\sqrt{m^2 c^2+\hbar^2 k_{0}^2}$ with spatial width $\sigma_z$. The state (\ref{elecstat}) is expressed in the interaction picture and constant shifts in the basis ($\ket{z+2 v_{0}t_{0}}\equiv\ket{z}$) can be ignored as they play no role in the time evolution of the observables considered here. We are primarily interested in quantum systems prepared in general superposition states. The initial state of the two-level emitter is thus given by
\be{
\ket{\psi_{qe}}=a\ket{g}+b \, e^{-i \phi_{r}}\ket{e}, \label{atomstat}
}
with $a^2+b^2=1$ and $\phi_{r}=\phi-\omega_{0}t_{0}$ being the relative phase of the coherent superposition. 

Since the interaction between the quantum emitter and the electron wavepacket is weak, the higher-order terms in the Magnus series can be neglected. We focus on the dynamics governed by the first term in the expansion, which reads
\bal{
\Omega_{1}=-\frac{i}{\hbar} \int^{t}_{t_{0}} H_{I}(t_{1}) dt_{1}, \label{omega}
}
where the interaction Hamiltonian $H_{I}(t)$ is given by
\bal{
H_{I}(t)=\big(\mf{d}_{eg}S^{+}e^{i\omega_{0}t}+\mf{d}_{eg}^{*} S^{-}e^{-i\omega_{0}t}\big)\cdot \vec{E}(r_{\perp},0,z+v_{0}t), \label{inthamil} 
}
and the electric field $\vec{E}(r_{\perp}, 0, z)$ of the propagating electron has the form
\bal{
\vec{E}(r_{\perp},0,z)=\frac{-e \gamma}{4 \pi \epsilon_{0}}\frac{(z \hat{e}_{z}+r_{\perp} \hat{e}_{x})}{(\gamma^{2}z^{2}+r_{\perp}^{2})^{3/2}}.\label{elecfield}
}
Substituting Eq.~(\ref{inthamil}) in Eq.~(\ref{omega}), we find
\bal{
\Omega_{1}= &-i \int^{t}_{t_{0}} dt_{1} \bigg(\frac{1}{\hbar} \big[\mf{d}_{eg}\cdot \vec{E}(r_{\perp},0,z+v_{0}t_{1})\big] S^{+}e^{i \omega_{0}t_{1}}\nonumber\\&+\textrm{H.c.}\bigg).
}
We now make a change of variables $s=z+v_{0}t_{1}$ and $ds=v_{0} dt_{1}$. The above integral can be written as
\bal{
\Omega_{1}= &\frac{-i}{\hbar v_{0}} \bigg( \left\{e^{-i \frac{\omega_{0}}{v_{0}}z}\!\!\!\int^{z+v_{0}t}_{z+v_{0}t_{0}}\!\!\! \!\,ds\, \big[\mf{d}_{eg}\cdot \vec{E}(r_{\perp},0,s)\big] e^{i \frac{\omega_{0} }{v_{0}}s}\right\} S^{+}  \nonumber\\ &~+\textrm{H.c.}\bigg).\label{integ}
}
We note that the relevant distance for the interaction would be a very small region, typically hundreds of nanometers around the location of the quantum emitter at the origin. In comparison, the electron traverses a much greater distance (on the order of hundreds of micrometers) from its source to the emitter's position, which allows us to extend the lower limit of the integral (\ref{integ}) to $-\infty$. Therefore, we have the following expression
\bal{
\Omega_{1}= &\frac{-i}{\hbar v_{0}} \bigg( \left\{e^{-i \frac{\omega_{0}}{v_{0}}z}\!\!\!\int^{z+v_{0}t}_{-\infty}\!\!\! \!\,ds\, \big[\mf{d}_{eg}\cdot \vec{E}(r_{\perp},0,s)\big] e^{i \frac{\omega_{0} }{v_{0}}s}\right\} S^{+}  \nonumber\\ &~+\textrm{H.c.}\bigg).\label{integ2}
}
In a more compact form, we can write Eq.~(\ref{integ2}) as
\bal{
\Omega_{1}= -i \left[\,g(z,t)\, S^{+}+\textrm{H.c.}\,\right],\label{omegfin}
}
where the \textit{time-dependent} coupling parameter reads
\bal{
g(z, t)=e^{-i \frac{\omega_{0}}{v_{0}}z}\left\{\eta \int^{z+v_{0}t}_{-\infty} \,ds\,\bigg[\frac{s\, \mf{d}_{z} +r_{\perp} \mf{d}_{x}}{(\gamma^{2}s^{2}+r_{\perp}^{2})^{3/2}}\bigg] e^{i \frac{\omega_{0}}{v_{0}} s}\right\}. \label{gpara}
}
In arriving at Eq.~(\ref{gpara}), we have made use of the electric field (\ref{elecfield}),  expressed the transition dipole moment in terms of its components $\mf{d}_{eg}=\mf{d}_{x} \hat{e}_{x}+\mf{d}_{y} \hat{e}_{y}+\mf{d}_{z} \hat{e}_{z}$, and set $\eta=-e \gamma/4\pi \epsilon_{0}\hbar v_{0}$. 

To proceed further, we employ the Leibniz rule to differentiate under the integral sign. Using this method, it can be shown that an integral $\mc{I}(\alpha)=\int^{\tilde{z}}_{-\infty}ds \,e^{i \alpha s}\,(\gamma^{2}s^{2}+r_{\perp}^2)^{-3/2}$ satisfies a nonhomogenous Bessel differential equation of the form $\mc{I}''(\alpha)-\alpha^{-1}\mc{I}'(\alpha)-\frac{r_{\perp}^2}{\gamma^{2}}\mc{I}(\alpha)=f(\alpha)$, where $\mc{I}'(\alpha)\equiv d\mc{I}/d\alpha$ and $\mc{I}''(\alpha)\equiv d^{2}\mc{I}/d\alpha^{2}$ are the first and second-order derivatives with respect to the parameter $\alpha$. It is then straightforward to find the solution $\mc{I}(\alpha)$ to this differential equation using standard techniques. 
We present here the final result, which is the exact semi-analytical solution for the time-dependent coupling parameter (\ref{gpara}), expressed as 
\bal{
g(z,t)=e^{-i \frac{\omega_{0}}{v_0}z}\big[g_{0}(z,t)+g_{1}(z,t)\big],\label{ganal}
}
with
\bal{
g_{0}=&~ \frac{g_{\infty}}{2} \left[1+\frac{\gamma (z+v_{0}t)}{\sqrt{\gamma^{2}(z+v_{0}t)^{2}+r_{\perp}^2}}\right],\nonumber\\ \\ 
g_{1}=&~\frac{\eta r_{\perp}\omega_{0}}{\gamma^{2}v_{0}\sqrt{\gamma^{2}(z+v_{0}t)^{2}+r_{\perp}^2}}\nonumber\\&\times  \bigg[g_{K}\!\int^{\frac{\omega_{0}}{v_{0}}}_{0}\!\! du \,\,\frac{e^{i u (z+v_{0}t)}}{u}I_{1}\left(\frac{r_{\perp}u}{\gamma}\right)\nonumber\\&-g_{I}\!\int^{\infty}_{\frac{\omega_{0}}{v_{0}}}\!\! du\,\, \frac{e^{i u (z+v_{0}t)}}{u}K_{1}\left(\frac{r_{\perp}u}{\gamma}\right)\bigg],\nonumber
}
The constant coefficients $g_{i}$ ($i=K,I$) read $g_{I}= [i \mf{d}_{x}I_{1}(r_{\perp}\omega_{0}/\gamma v_{0})+\mf{d}_{z}I_{0}(r_{\perp}\omega_{0}/\gamma v_{0}))/\gamma]$ and $g_{K}=[i \mf{d}_{x}K_{1}(r_{\perp}\omega_{0}/\gamma v_{0})+\mf{d}_{z}K_{0}(r_{\perp}\omega_{0}/\gamma v_{0}))/\gamma]$, where $I_{v}(x)$ and $K_{v}(x)$ are the modified Bessel functions of the first and second kind, respectively. It is easy to see that in the long-time limit $t \rightarrow \infty$, $g_{1} \rightarrow 0$ and $g_{0}(z,t)$ reduces to 
\bal{
\lim_{t \rightarrow \infty} g_{0}(z,t)\equiv g_{\infty}=&\frac{2\eta \omega_{0}}{\gamma^{2} v_{0}} \bigg\{\mf{d}_{x} K_{1}\left(\frac{r_{\perp}\omega_{0}}{\gamma v_{0}}\right)\nonumber\\&+\frac{i \mf{d}_{z}}{\gamma} K_{0}\left(\frac{r_{\perp}\omega_{0}}{\gamma v_{0}}\right)\bigg\},\label{ginf}
}
which is the standard coupling constant that arises in a full scattering matrix description of a free electron interacting with a two-level system \cite{Ruimy2021, Zhao2021}. Interestingly, the parameters $g_{i}(z,t)$ ($i=0,1$) depend on the variables $z$ and $t$ as $\tilde{z}=z+v_{0}t$. Taking advantage of this fact, it is useful to go to the spectral domain. Using the Fourier convention, $g(x)=\frac{1}{2\pi}\int \! d\xi  \, f(\xi)\, e^{-i \xi x}$ and $f(\xi)=\int \! dx\,g(x) e^{i \xi x}$, we find 
\be{
G(s)=g_{0}(s)+g_{1}(s)=\frac{1}{2\pi} \int_{-\infty}^{\infty} \!d\xi\, F(\xi) e^{-i s \xi},\label{g1}
}
where $F(\xi)$ is given by
\bal{
F(\xi)=  \pi \delta(\xi) \,g_{\infty}+ g_{c}(\xi), \label{Fomega}
}
and
\bal{
g_{c}(\xi)=& \frac{i r_{\perp}}{\gamma}\bigg[g_{\infty}\, K_{1}\left(\frac{r_{\perp}|\xi|}{\gamma }\right)\mathrm{sgn}(\xi)\bigg]\nonumber\\&+\frac{2 \eta r_{\perp}\omega_{0}}{\gamma^{2}v_{0}}\bigg[g_{K}\!\!\int^{\frac{\omega_{0}}{v_{0}}}_{0}\!\!\! du \,\,\frac{1}{u} I_{1}\!\!\left(\frac{r_{\perp}u}{\gamma}\right)K_{0}\!\left(\frac{r_{\perp}|u +\xi|}{\gamma}\right)\nonumber\\&-g_{I}\!\!\int^{\infty}_{\frac{\omega_{0}}{v_{0}}}\!\!\! du\,\, \frac{1}{u}K_{1}\!\!\left(\frac{r_{\perp}u}{\gamma}\right)K_{0}\!\left(\frac{r_{\perp}|u+\xi|}{\gamma}\right)\!\!\bigg], \label{gcomega}
} 
with
\be{
\operatorname{sgn}(\xi)
=
\begin{cases}
+1, & \xi > 0,\\[2pt]
0,  & \xi = 0,\\[2pt]
-1, & \xi < 0.\nonumber
\end{cases}
} 
Setting $s=z+v_{0}t$ and $\xi=\omega/v_{0}$ in the above expressions, the coupling parameter in Eq.~(\ref{ganal}) can be written as
\bal{
g(z,t)=&~e^{-i \frac{\omega_{0}}{v_{0}}z}G(z+v_{0}t)\nonumber\\=&~\frac{1}{2\pi v_{0}} \int \!\!d\omega\, F\bigg(\frac{\omega-\omega_{0}}{v_{0}}\bigg) e^{-i \frac{\omega}{v_{0}}z}e^{-i (\omega-\omega_{0}) t}. \label{gspecf}
}
Here, we identify the electron momentum ladder operators as
\be{
b(\omega) =e^{-i \frac{\omega}{v_{0}}z},~~~~ b^{\dagger}(\omega) =e^{i \frac{\omega}{v_{0}}z},
}
where $b(\omega)$ [$b^{\dagger}(\omega)$] lowers [raises] the electron momentum by $\hbar \omega/v_{0}$. We now see that the coupling parameter $g(z,t)$ can be viewed as the electron ladder operator $b(\omega)$ weighted by the function $F([\omega-\omega_{0}]/v_{0})$ given by Eq.~($\ref{Fomega}$). The Dirac delta $\delta(\omega-\omega_{0})$ is therefore responsible for the free electron gaining or losing exactly the energy $\hbar \omega_{0}$ of the two-level transition with a constant coefficient $g_{\infty}$. In contrast, the spectral weight $g_{c}(\omega)$ evolves as $e^{-i (\omega-\omega_{0})t}$ and leads to changes in the electron energy that is detuned from the transition energy $\hbar \omega_{0}$. 

In Appendix B and C, we make use of the solution (\ref{gspecf}) to understand the qualitative features in the population dynamics and the EELS spectra.  

Finally, with the dynamics governed only by $\Omega_{1}$ (\ref{omegfin}), it is straightforward to express the solution (\ref{sol}) in the position basis $\int dz \ket{z}\bra{z}$ and write the joint quantum state of the electron wavepacket and the two-level emitter as
\be{
\ket{\psi(t)} = \int dz\, e^{-\frac{z^2}{4\sigma_{z}^2}+i k_{0} z} [c_{g}(z,t) \ket{g}+c_{e}(z,t)\ket{e}]\ket{z},
}
where the initial states of the quantum emitter and the free electron are according to Eqs.~(\ref{elecstat}) and (\ref{atomstat}). The probability amplitudes are found to be
\bal{
c_{g}(z,t)=&~\big[\,a \cos{|g(z,t)|}-i\, b \,e^{-i( \phi_{r}+\phi_{g}(z,t))}\sin{|g(z,t)|}\,\big],&\nonumber\\
c_{e}(z,t)=&~\big[\,b\, e^{-i \phi_{r}} \cos{|g(z,t)|}-i\, a\, e^{i\phi_{g}(z,t)}\sin{|g(z,t)|}\,\big].\nonumber\\\label{probamp2}
} 
\section{Population Dynamics}
In this section, we derive a semianalytical form for the populations in the case of general superposition states of the quantum emitter. The qualitative features of our results can be easily understood with the help of these solutions. To investigate the quantum dynamics of the two-level emitter, we compute the reduced density matrix
\bal{
\rho_{qe}(t)=&~\textrm{Tr}_{el}\{\ket{\psi(t)}\bra{\psi(t)}\}=\frac{1}{\sqrt{2 \pi}\sigma_{z}}\!\!\int dz \, e^{- \frac{z^2}{2\sigma_z^2}}\nonumber\\&\times 
\begin{pmatrix}
|c_g(z,t)|^2 & c_g(z,t)\, c_e^*(z,t) \\
c_e(z,t)\, c_g^*(z,t) & |c_e(z,t)|^2
\end{pmatrix}, \label{dmat}
}
where the diagonal elements $|c_{g}|^2$ and $|c_{e}|^2$ give the populations of the ground and excited states, respectively. Under the weak coupling assumption $|g(z,t)|\ll1$, we can write the inversion $\rho_{ee}-\rho_{gg}$ approximately as
\bal{
\rho_{ee}-\rho_{gg}=&~\frac{1}{\sqrt{2 \pi}\sigma_{z}}\int dz \,e^{-\frac{z^2}{2 \sigma_{z}^2}} \big[b^2-a^2\nonumber\\&+4 \,a b \, \textrm{Im}\{g(z,t) e^{i \phi_{r}}\}\big],\label{inv}
}
whereas the trace $\rho_{ee}+\rho_{gg}$ of the density matrix (\ref{dmat}) naturally satisfies
\bal{
\rho_{ee}+\rho_{gg}=&~\frac{1}{\sqrt{2 \pi}\sigma_{z}}\int dz\, e^{-\frac{z^2}{2 \sigma_{z}^2}} (a^2+b^2)=1.
}
The expression (\ref{inv}) can be further simplified to yield
\bal{
\rho_{ee}-\rho_{gg}= &~b^2-a^2+2\,ab\,e^{-\frac{\sigma_{z}^2 \omega_{0}^{2}}{2v_{0}^2}}\textrm{Im}\big\{g_\infty \,e^{i\phi_{r}}\big\}\nonumber\\&+ \,\frac{2ab}{v_{0}}\,   \textrm{Im}\bigg\{\!\!\int d\omega \,g_{c}\bigg(\frac{\omega-\omega_{0}}{v_{0}}\bigg)\,e^{-\frac{\sigma_{z}^2 \omega^{2}}{2v_{0}^2}} \nonumber\\& \times e^{-i (\omega-\omega_{0})t+i\phi_{r}}\bigg\},\nonumber \\
}
where we have made use of Eq.~(\ref{gspecf}). We can therefore write the equations for the excited and ground state populations as
\bal{
\rho_{ee}=&~b^2+\,ab\,e^{-\frac{\sigma_{z}^2 \omega_{0}^{2}}{2v_{0}^2}}\textrm{Im}\big\{g_\infty \,e^{i\phi_{r}}\big\}\nonumber\\&+ \,\frac{ab}{v_{0}}\,   \textrm{Im}\bigg\{\!\!\int d\omega \,g_{c}\bigg(\frac{\omega-\omega_{0}}{v_{0}}\bigg)\,e^{-\frac{\sigma_{z}^2 \omega^{2}}{2v_{0}^2}} \nonumber\\& \times e^{-i (\omega-\omega_{0})t+i\phi_{r}}\bigg\},\nonumber\\
\rho_{gg}=&~a^2-\,ab\,e^{-\frac{\sigma_{z}^2 \omega_{0}^{2}}{2v_{0}^2}}\textrm{Im}\big\{g_\infty \,e^{i\phi_{r}}\big\}\nonumber\\&- \,\frac{ab}{v_{0}}\,   \textrm{Im}\bigg\{\!\!\int d\omega \,g_{c}\bigg(\frac{\omega-\omega_{0}}{v_{0}}\bigg)\,e^{-\frac{\sigma_{z}^2 \omega^{2}}{2v_{0}^2}} \nonumber\\& \times e^{-i (\omega-\omega_{0})t+i\phi_{r}}\bigg\}.\label{pop}
}
It is obvious that the dynamics is entirely governed by the spectral coupling parameter $g_{c}(\omega)$. In addition, we can clearly see why the coherent oscillations cannot occur for an initial ground ($a=1$, $b=0$) or excited state ($a=0$, $b=1$) of the two-level emitter. In such cases, the approximate expressions (\ref{pop}) show that the populations are nearly unchanged for weak interactions. However, for an initial superposition state, interesting coherence effects are induced by the electron wavepacket. A numerical analysis reveals that $g_{c}(\omega)$ is dominant around $\omega = \omega_{0}$ whereas the Gaussian prefactor $e^{-\sigma_{z}^2 \omega^2/(2 v_{0}^2)}$ has the maximum weight near $\omega=0$. Therefore, heuristically considering $g_{c}(\omega)\approx g_{c}(\omega=0)$ and pulling the spectral weight outside the integral, we see that the populations behave according to
\bal{
\rho_{ee}\sim &~b^2+\,ab\,e^{-\frac{\sigma_{z}^2 \omega_{0}^{2}}{2v_{0}^2}}\textrm{Im}\big\{g_\infty \,e^{i\phi_{r}}\big\}\nonumber\\&+\sqrt{2\pi}  \,ab\,   \frac{e^{-\frac{t^2}{2\sigma_{t}^2}}}{\sigma_{z}}\textrm{Im}\bigg\{g_{c}\bigg(\frac{-\omega_{0}}{v_{0}}\bigg)e^{i (\omega_{0}t+\phi_{r})}\bigg\},\nonumber\\
\rho_{gg}\sim &~a^2-\,ab\,e^{-\frac{\sigma_{z}^2 \omega_{0}^{2}}{2v_{0}^2}}\textrm{Im}\big\{g_\infty \,e^{i\phi_{r}}\big\}\nonumber\\&- \sqrt{2\pi}  \,ab\,   \frac{e^{-\frac{t^2}{2\sigma_{t}^2}}}{\sigma_{z}}\textrm{Im}\bigg\{g_{c}\bigg(\frac{-\omega_{0}}{v_{0}}\bigg) e^{i (\omega_{0}t+\phi_{r})}\bigg\}, \label{popbeh}
}
where $\sigma_{t}=\sigma_{z}/v_{0}$ is the duration of the wavepacket. The coherent oscillations therefore arise from the exponential oscillatory term $e^{i \omega_{0}t}$ with a temporal Gaussian envelope $e^{-t^2/(2\sigma_{t}^2)}$. Thus, we have qualitatively shown the origin of the induced coherence effects in the populations.
\section{Quantum coherence effects in the EELS spectra}
In this section, we discuss the electron energy spectra and show how the coherent features arise in the zero-loss peak of the spectral probability density. Let us consider the probability difference, 
\bal{
\Gamma_{net}(E,t)=\frac{1}{\hbar v_{0}}\big(|\psi_{e}(k(E),t)|^2-|\psi_{g}(k(E),t)|^2 \big),
}
where
\bal{
\psi_{i}(k(E),t)= \frac{1}{N}\!\int \! dz \,e^{-z^2/4 \sigma_{z}^2}e^{-i k z}c_{i}(z,t).\label{eelsprob}
}
In this expression, $N=\sqrt{2\pi}\,(2\pi\sigma_z)^{1/4}$ is a multiplicative factor, $k(E)=(E-E_{0})/\hbar v_{0}$, and $c_{i}(z,t)$ ($i=g,e$) are given by Eq.~(\ref{probamp}). In the regime of weak interactions $g(z,t)\ll1$, we can write Eq.~(\ref{eelsprob}) as
\bal{
\psi_{g}=&~\frac{1}{N}\!\int \! dz \,e^{-z^2/4 \sigma_{z}^2}e^{-i k z}[a-i\, b\, e^{-i \phi_{r}}g^{*}(z,t)],\nonumber\\
\psi_{e}=&~\frac{1}{N}\!\int \! dz \,e^{-z^2/4 \sigma_{z}^2}e^{-i k z}[b\,e^{-i \phi_{r}}-i\, a\, g(z,t)].
}
For brevity, we adopt the short-hand notation $\psi_{i}\equiv \psi_{i}(k(E),t)$ ($i=g,e$). We now proceed in a manner similar to that in the previous analysis. Once again employing Eq.~(\ref{gspecf}), the amplitudes $\psi_{i}$ can be expressed as
\bal{
\psi_{g}=&~\frac{2\sqrt{\pi}\sigma_{z}}{N}\bigg[a\, e^{-k^2 \sigma_{z}^2}-\frac{i\,b\,e^{-i \phi_{r}}}{2}g_{\infty}^{*}e^{-\sigma_{t}^2 v_{0}^2\left(k-\frac{\omega_{0}}{v_{0}}\right)^2}\nonumber\\&-\frac{i\,b \,e^{-i \phi_{r}}}{2\pi v_{0}}\!\!\!\int \!\! d\omega\, g_c^{*}\!\bigg(\!\frac{\omega-\omega_{0}}{v_{0}}\!\bigg)e^{i (\omega-\omega_{0}) t} e^{-\sigma_{t}^2 v_{0}^2\left(\!k-\frac{\omega}{v_{0}}\!\right)^2}\bigg],\nonumber\\ \nonumber\\
\psi_{e}=&~\frac{2\sqrt{\pi}\sigma_{z}}{N}\bigg[b \,e^{-i \phi_{r}} e^{-k^2 \sigma_{z}^2}-\frac{i\,a}{2}\,g_{\infty}\,e^{-\sigma_{t}^2 v_{0}^2\left(k+\frac{\omega_{0}}{v_{0}}\right)^2}\nonumber\\&-\frac{i\,a}{2\pi v_{0}}\!\int \!\! d\omega\, g_c\!\bigg(\!\frac{\omega-\omega_{0}}{v_{0}}\!\bigg)e^{-i (\omega-\omega_{0}) t} e^{-\sigma_{t}^2 v_{0}^2\left(\!k+\frac{\omega}{v_{0}}\!\right)^2}\bigg],
}
where $g_c^{*}(\omega)=[g_c(\omega)]^{*}$ denotes the complex conjugate. It is clear that the second term in $\psi_{i}$ gives rise to the sidebands in the EELS spectra leading to electron energy gain or loss of $\hbar \omega_{0}$ whereas the interesting oscillatory features originate from the spectral weight $g_{c}(\omega)$. To simplify the calculations, we consider an equally weighted superposition $a=b$, as in the main text. Since we are interested in the zero-loss peak, we set $k(E)=(E-E_{0})/\hbar v_{0}=0$ in the above expression and assume that $g_{c}(\omega)\approx g_{c}(\omega=0)$ since the Gaussian prefactor is centered at $\omega=0$, which is far from the resonance of $g_{c}(\omega=\omega_{0})$. Therefore, the time evolution of the amplitudes behaves according to
\bal{
\psi_{g}=&~\frac{2\sqrt{\pi}\sigma_{z}}{N}\bigg[a-\frac{i\,a\,e^{-i \phi_{r}}}{2}g_{\infty}^{*}e^{-\sigma_{t}^2 \omega_{0}^2}\nonumber\\&-\frac{i\,a}{2\sqrt{\pi} \sigma_{z}} g_c^{*}\!\bigg(\!\frac{-\omega_{0}}{v_{0}}\!\bigg)e^{-i (\omega_{0} t+\phi_{r})} e^{-\frac{t^2}{4 \sigma_{t}^2}}\bigg],\nonumber\\ \nonumber\\
\psi_{e}=&~\frac{2\sqrt{\pi}\sigma_{z}\,e^{-i \phi_{r}}}{N}\bigg[a -\frac{i\,a\,e^{i \phi_{r}}}{2}\,g_{\infty}\,e^{-\sigma_{t}^2 \omega_{0}^2}\nonumber\\&-\frac{i\,a}{2\sqrt{\pi} \sigma_{z}}g_c\!\bigg(\!\frac{-\omega_{0}}{v_{0}}\!\bigg)e^{i (\omega_{0}t+\phi_{r})}e^{-\frac{t^2}{4 \sigma_{t}^2}} \bigg],
}
Using these relations, we find the probability difference $\Gamma_{net}(E,t)$ to be
\bal{
\Gamma_{net}(E,t)\sim &\frac{1}{\hbar v_{0}}\left(\frac{8\pi a^2 \,\sigma_{z}^2}{N^2}\right) \bigg[e^{-\sigma_{t}^2\omega_{0}^2}\textrm{Im}\big\{g_{\infty}e^{i\phi_{r}}\big\}\nonumber\\&+\frac{e^\frac{-t^2}{4\sigma_{t}^2}}{\sqrt{\pi}\sigma_{z}}\textrm{Im}\bigg\{g_c\!\bigg(\!\frac{-\omega_{0}}{v_{0}}\!\bigg)e^{i(\omega_{0}t+\phi_{r})}\bigg\} \bigg].
}
This expression is identical to the equations (\ref{popbeh}) describing the behaviour of the populations. We find that the zero-loss peak coherently oscillates at the transition frequency $\omega_{0}$ of the two-level emitter with a broad Gaussian envelope $e^{-t^2/(4 \sigma_{t}^2)}$. Thus, a signature of the quantum coherence of the emitter is found in the EELS dynamics even for unmodulated electron wavepackets.

\section{Quantum coherence effects in the cathodoluminescence emission}
In this section, we apply our theoretical framework to derive an analytical expression for the cathodoluminescence (CL) spectrum. Furthermore, we show that the predicted coherence effects can be observed in ultrafast spectroscopic CL based on electron-driven photon sources. We begin with the general definition of the physical spectrum introduced by Eberly and W\'odkiewicz \cite{Eberly1977}. This formalism generalizes the conventional steady-state spectrum by accounting for the filtering of the light field prior to detection, for example, by a Fabry--P\'erot interferometer. In general, the spectrum can be written as
\bal{
S(\omega, \Gamma; t)=&~\Gamma \int_{t_0}^t dt_1 \int_{t_{0}}^t dt_{2}\, e^{-(\frac{\Gamma}{2}-i \omega)(t-t_{1})}e^{-(\frac{\Gamma}{2}+i \omega)(t-t_{2})} \nonumber\\  &\times \langle \vec{E}^{(-)}(r,t_1) \cdot \vec{E}^{(+)}(r,t_2)\rangle, \label{spectra}
}
where $\omega$ is the observed frequency, $\Gamma$ denotes the spectral bandwidth of the filter, and $t_0$ is the initial time. For a two-level system, the positive frequency part of the electric-field operator \cite{Cohen1998} in the far-field at the position of the detector $\vec{r}=r \hat{r}$ can be written as 
\bal{
\vec{E}^{(+)}(r,t)= f(r)\, S^{-}(t-r/c) \, \hat{e},
}
where $f(r)=\omega_{0}^2 |\mf{d}_{eg}|/(4 \pi \epsilon_{0}c^2 r)$ is a multiplicative factor, $S^{-}$ is the lowering operator, and $\hat{e}=\hat{e}_d-\hat{r}\,(\hat{r}\cdot \hat{e}_d)$ denotes the polarization direction of the field with $\hat{e}_d=\mf{d}_{eg}/|\mf{d}_{eg}|$ being the unit vector in the direction of the dipole moment. Substituting the above expression for the electric-field operator in Eq.~(\ref{spectra}), we find
\bal{
S(\omega)=&~\int_{t_0}^t dt_1 \int_{t_{0}}^t dt_{2}\, e^{-(\frac{\Gamma}{2}-i \omega)(t-t_{1})}e^{-(\frac{\Gamma}{2}+i \omega)(t-t_{2})} \nonumber\\  &\times \langle S^{+}(t_1-r/c) S^{-}(t_2-r/c)\rangle, \label{spectra2}
}
where we have scaled $S(\omega)\equiv S(\omega)/(\Gamma f^2(r))$ with respect to the factor $f(r)$ and adopt the short-hand notation $S(\omega)\equiv S(\omega, \Gamma; t)$. 

We focus on the coherent component of the spectrum in Eq.~(\ref{spectra2}). To this end, we decompose the transition operators into their mean values and fluctuations \cite{Cohen1998} according to
\[
\delta S^{\pm}(t)
=
S^{\pm}(t)-\langle S^{\pm}(t)\rangle .
\]
This allows the spectrum in Eq.~(\ref{spectra2}) to be expressed as
\be{
S(\omega)
=
S_{\textrm{coh}}(\omega)
+
S_{\textrm{incoh}}(\omega),
}
where $S_{\textrm{coh}}(\omega)$ and $S_{\textrm{incoh}}(\omega)$ denote the coherent and incoherent contributions to the CL spectrum, respectively. The experimentally relevant observable is the coherent part of the spectrum, defined by $S_{CL}(\omega)
=
S(\omega)-S_{\textrm{incoh}}(\omega)
=
S_{\textrm{coh}}(\omega)$. The resulting spectral response in the long-time limit is then determined by the induced coherence and can be written as 
\be{
S_{CL}(\tilde{\omega}, \tau)=\left|\int_{\tau}^{\infty}\!\!dt\, e^{i \tilde{\omega} t} \rho_{eg}(t,\tau)\right|^2, \label{specform}
}
where the spectrum is expressed as a function of the detuning $\tilde{\omega}=\omega-\omega_{0}$ from the transition frequency $\omega_{0}$, and of the electron probe arrival time $\tau$. The filter-induced broadening is neglected under the assumption $\Gamma \ll \gamma$. In the above expression, $\rho_{eg}(t,\tau)$ denotes the induced coherence in the presence of decay and dephasing in the interaction picture defined by $U=\exp\{i H_{0} t/\hbar\}$. 

To properly account for decoherence effects, we first derive the equation of motion for the off-diagonal density matrix element $\rho_{eg}(t)$ of the coupled quantum emitter–free-electron system. Dissipative processes can then be incorporated through the Lindblad master-equation formalism \cite{Breuer2009, Crispin2025}. The dynamics of the combined state of the quantum emitter and the free electron are governed by the Liouville-Von Neumann equation
\be{
\frac{\partial \rho}{\partial t}=-\frac{i}{\hbar}[H_{I}(t),\rho(t)], \label{maindmeq}
}
where $\rho(t)$ is the density operator in the interaction picture and $H_{I}(t)=U(H-H_{0})U^{\dagger}$ is the interaction Hamiltonian given by Eq.~(\ref{inthamil}). Since we are interested in the evolution of the coherence, we first trace over the electron degrees of freedom in Eq.~(\ref{maindmeq}) and then evaluate the dynamics of the density-matrix element $\rho_{eg}\equiv \matelal{e}{\rho_{qe}}{g}$, where $\rho_{qe}=\textrm{Tr}_{\textrm{el}}\{\rho\}$ denotes the reduced density matrix of the quantum emitter. Following this procedure and using Eq.~(\ref{inthamil}), we obtain the equation of motion for the coherence:
\bal{
 \Dot{\rho}_{eg}=&-\frac{i}{\hbar}\int \!\! dz\, e^{-\frac{z^2}{2 \sigma_{z}^2}} e^{i \omega_{0}t} (\mf{d}_{eg}\cdot \vec{E}(z+v_{0}t))\nonumber\\&\times  \big[|c_{g}(z,t)|^2-|c_{e}(z,t)|^2\big]. \label{reducedeq}
}
Equation~(\ref{reducedeq}) is obtained by substituting the solution from Eq.~(\ref{main}) into the right-hand side of the Liouville equation~(\ref{maindmeq}). Substituting the probability amplitudes from Eq.~(\ref{probamp}) into Eq.~(\ref{reducedeq}), we then obtain
\bal{
 \Dot{\rho}_{eg}=&-\frac{i}{\hbar}\int \!\! dz\, e^{-\frac{z^2}{2 \sigma_{z}^2}} e^{i \omega_{0}t} (\mf{d}_{eg}\cdot \vec{E}(z+v_{0}t))\nonumber\\& \times  \big[(a^2-b^2)\cos{2|g(z,t)|}\nonumber\\ &-2\, a \,b \sin{2|g(z,t)|}\sin{(\phi_{r}+\phi_{g}(z,t))}\big]. \label{coherent}
}
We now include the damping terms describing radiative decay and dephasing into the coherent contribution in Eq.~(\ref{coherent}). In the interaction picture, defined by $U=\exp(i H_{0}t/\hbar)$, the time evolution of the coherence can be written as
\be{
 \Dot{\rho}_{eg}=-\frac{1}{2} (\gamma_r+2\gamma_{nr})\rho_{eg}+f_{el}(t),
 \label{coheq}
}
where $\gamma_{r}$ denotes the radiative decay rate and $\gamma_{nr}$ represents the pure dephasing rate of the quantum emitter, arising, for example, from phonons and other nonradiative relaxation channels. The electron-induced coherence term $f_{el}(t)$ is given by
\bal{
f_{el}(t)=&-\frac{i}{\hbar}\int \!\! dz\, e^{-\frac{z^2}{2 \sigma_{z}^2}} e^{i \omega_{0}t} (\mf{d}_{eg}\cdot \vec{E}(z+v_{0}t))\nonumber\\& \times  \big[(a^2-b^2)\cos{2|g(z,t)|}\nonumber\\ &-2\, a \,b \sin{2|g(z,t)|}\sin{(\phi_{r}+\phi_{g}(z,t))}\big].
}
To proceed further, we approximate $\sin(2|g(z,t)|)\simeq 2g(z,t)$ and $\cos(2|g(z,t)|)\simeq 1$, which is justified in the weak-coupling regime $|g(z,t)|\ll 1$. Under this approximation, the above equation reduces to
\bal{
 f_{el}(t)=&~-\frac{i}{\hbar}(a^2-b^2)\int \!\! dz\, e^{-\frac{z^2}{2 \sigma_{z}^2}} e^{i \omega_{0}t} (\mf{d}_{eg}\cdot \vec{E}(z+v_{0}t))\nonumber\\&~+\frac{2 a b}{\hbar}\!\!\int \!\! dz\, e^{-\frac{z^2}{2 \sigma_{z}^2}} e^{i \omega_{0}t} (\mf{d}_{eg}\cdot \vec{E}(z+v_{0}t))\nonumber\\& \times \big[g(z,t)e^{i \phi_{r}}-g^{*}(z,t)e^{-i \phi_{r}}\big].
}
The advantage of the above expression is that the trace integral $(\int dz)$ can now be evaluated using the Fourier transforms in Eqs.~(\ref{g1})--(\ref{gcomega}) for the coupling parameter $g(z,t)$ and the electric field $\vec{E}(z+v_{0}t)$. To this end, we rewrite the integral after the exchange of variables $s=z+v_{0}t$ and using Eq.~($\ref{ganal}$), 
\bal{
f_{el}(t)=&~-\frac{i}{\hbar}(a^2-b^2)e^{i \omega_{0}t} \int \!\! ds\, e^{-\frac{(s-v_{0}t)^2}{2 \sigma_{z}^2}} (\mf{d}_{eg}\cdot \vec{E}(s))\nonumber\\&~+\frac{2ab}{\hbar} e^{i \omega_{0}t}\!\! \int ds \,e^{-\frac{(s-v_{0}t)^2}{2 \sigma_{z}^2}}  (\mf{d}_{eg}\cdot \vec{E}(s))\nonumber\\ &\times \big[(g_{0}(s)+g_{1}(s))e^{i \phi_{r}}e^{-i \frac{\omega_{0}}{v_{0}}s} e^{i \omega_{0}t}-c.c.\big].\label{fel}
}
Using the Fourier convention, $g(x)=\frac{1}{2\pi}\int \! d\xi  \, f(\xi)\, e^{-i \xi x}$ and $f(\xi)=\int \! dx\,g(x) e^{i \xi x}$, we can express $\mf{d}_{eg}\cdot \vec{E}(s)$ as 
\be{
\mc{E}(s) =\mf{d}_{eg}\cdot \vec{E}(s)=\frac{1}{2\pi}\int_{-\infty}^{\infty} \! d\xi \,\mc{E}(\xi) e^{-i \xi s},\label{e1}
}
with
\be{
\mc{E}(\xi)=\frac{-e}{2\pi \gamma \epsilon_{0}}\bigg[\mf{d}_x \,|\xi|\, K_{1}\left( \frac{r_{\perp}|\xi|}{\gamma}\right)+i \frac{\mf{d}_z}{\gamma}\, \xi \,K_{0}\left( \frac{r_{\perp}|\xi|}{\gamma}\right) \bigg]. \\ \vs
}
With the help of Eqs.~(\ref{g1}) and (\ref{e1}), the resulting integral ($\int ds $) in Eqs.~(\ref{fel}) yields
\bal{
f_{el}(t)=&~\frac{(a^2-b^2)}{\hbar v_{0}}\int \!d\omega\,f_{0}(\omega)\,e^{-i (\omega-\omega_{0}) t}\nonumber\\&+\frac{ab}{\pi \hbar v_{0}} \, \int \!\! d\omega\, f_{el}(\omega) \, e^{-i (\omega-\omega_{0}) t}, \label{specresp}
}
where
\bal{
f_{0}(\omega)&=\frac{-i \sigma_{z}}{\sqrt{2\pi}}\, \mc{E}\!\left(\frac{\omega}{v_{0}}\right)e^{-\frac{\sigma_{z}^2\omega^2}{2v_{0}^2}},\nonumber\\ \nonumber\\ 
f_{el}(\omega)&=\nonumber\\&~\frac{\sigma_{z}}{\sqrt{2\pi}} \bigg\{\!\bigg[\pi g_{\infty} \mc{E}\!\left(\frac{\omega}{v_{0}}\right)\!+\!(\mc{E}*g_{c})\!\left(\frac{\omega}{v_{0}}\right)\!\bigg]\!e^{i \phi_{r}-\frac{\sigma_{z}^2(\omega+\omega_{0})^2}{2v_{0}^2}}\nonumber\\&-\!\bigg[\pi g^{*}_{\infty} \mc{E}\!\left(\frac{\omega}{v_{0}}\right)\!+\!(\mc{E}*\tilde{g}_{c})\!\left(\frac{\omega}{v_{0}}\right)\!\bigg]e^{-i \phi_{r}-\frac{\sigma_{z}^2(\omega-\omega_{0})^2}{2v_{0}^2}}\bigg\}.\nonumber\\
}
Here, $(f*g)(\omega)=\int_{-\infty}^{\infty} \! d\xi \,f(\xi) g(\omega-\xi) $ denotes the convolution and $\tilde{g}_{c}(\omega)=g_{c}^{*}(-\omega)$. We now return to Eq.~(\ref{coheq}), whose solution can be written as
\be{
\rho_{eg}(t)=e^{-\gamma_{d}(t-t_{0})}\rho_{eg}(t_{0})+ \int_{t_{0}}^{t} dt' e^{-\gamma_{d}(t-t')}f_{el}(t'),
}
where $\gamma_{d}=(\gamma_{r}+2\gamma_{nr})/2$ denotes the total decoherence rate, and $t_{0}$ is the preparation time of the initial coherent superposition state. The above expression is valid in the weak electron--emitter coupling regime for arbitrary initial coherent superposition states of the quantum emitter.

In ultrafast CL spectroscopic studies, the characteristic timescales associated with radiative decay and dephasing are typically much longer than the interaction time between the free electron and the quantum emitter. Consequently, the electron--emitter interaction can be treated as effectively instantaneous, i.e.,
$f_{el}(t') \rightarrow f_{el}(t')\,\delta(t'-\tau)$ $(\tau \ge t_{0})$. Therefore, we have the solution
\be{
\rho_{eg}(t, \tau)=e^{-\gamma_{d}(t-t_{0})}+ e^{-\gamma_{d}(t-\tau)}f_{el}(\tau), \label{rhoeg}
}
where $\rho_{eg}(t,\tau)$ is the coherence as a function of the electron probe arrival time $\tau$. 

As in the main text, we consider an equally weighted coherent superposition state, $\rho_{eg}(t_{0})=a \,b \,e^{-i \phi_{r}}=a^2 e^{-i \phi_{r}}$. In this case, only the second term in Eq.~(\ref{specresp}) contributes. Finally, using Eqs.~(\ref{rhoeg}) and (\ref{specresp}), we obtain a compact analytical form for the spectra (\ref{specform}) given by
\bal{
S_{CL}(\tilde{\omega}, \tau)=&\bigg|\frac{e^{-\gamma_{d}(\tau-t_{0})}
}{\gamma_{d}-i\, \tilde{\omega}} \rho_{eg}(t_{0})\nonumber\\&+\frac{a b}{\pi \hbar v_{0}} \int \!\!d\Omega \,\frac{f_{el}(\Omega+\omega_{0})\, e^{-i\Omega\tau}}{\gamma_{d}-i \tilde{\omega}}\bigg|^2,
}
where $\tilde{\omega}=\omega-\omega_{0}$ is the detuning of the observed frequency $\omega$ from the transition frequency $\omega_{0}$ and $\tau$ denotes the arrival time of the electron probe. The quantity $S_{CL}(\tilde{\omega},\tau)$ represents the coherent component of the CL radiation and depends on the coherence $\rho_{eg}(t,\tau)$ induced by the free electron.



\bibliography{edphs}
\end{document}